# On quantum neural networks

**Alexandr A. Ezhov**

*State Research Center of Russian Federation*
*Troitsk institute for innovation and fusion research,*
*Moscow, Russia, 108840*
ezhov@triniti.ru

The early definition of a quantum neural network as a new field that combines the classical neurocomputing with quantum computing was rather vague and satisfactory in the 2000s. The widespread in 2020 modern definition of a quantum neural network as a model or machine learning algorithm that combines the functions of quantum computing with artificial neural networks [1] deprives quantum neural networks of their fundamental importance. We argue that the concept of a quantum neural network should be defined in terms of its most general function as a tool for representing the amplitude of an arbitrary quantum process. Our reasoning is based on the use of the Feynman path integral formulation in quantum mechanics. This approach has been used in many works to investigate the main problem of quantum cosmology, such as the origin of the Universe (see, for example, [2]). In fact, the question of whether our Universe is a quantum computer was posed by Seth Lloyd [3], who gave the answer is "yes", but we argue that the universe can be thought of as a quantum neural network.

## 1. Introduction

Last year, the results of a scientometric analysis of research in the field of "Quantum Neural Networks" (QNN) were [1]. The data for the study were taken from the Scopus database for the period 1990-2019 (the terms "quantum" and "neural networks" were used in the Scopus search engine), and the authors of [1] found 456 publications. The definition of a quantum neural network used by the authors − "*machine learning model or algorithm that combines the features of quantum computing with artificial neural networks*" − has been attributed to articles by Subhash Kak [4] and one of the authors and Dan Ventura [5]. It is not so important that Subhash Kak did not use the term "quantum neural network" (in fact, he introduced the concepts of quantum neural computing as well as quantum neural computer) in his groundbreaking paper. What's even more intriguing is that the authors of both papers [4,5] never used the term "machine learning". Currently, classical neural networks are often viewed as part of machine learning which, on the other hand, as only one of the parts of artificial intelligence (AI) [6]. Here we will try to show that this is generally not true. Moreover, we also trying to argue that intelligence, whether natural or artificial, as well as machine learning, together with the specialists working in these and other fields of science, can be considered as parts of a kind of quantum neural network, because the Universe in with we live, can also be thought of as a global quantum neural network. Let's start with the definition of a quantum neural computer given by Subhash Kak [4]:

> "*We define a quantum neural computer as a strongly connectionist system that is nevertheless characterized by a wavefunction*" and also "*In contrast to a quantum computer, which consists of quantum gates as components, a quantum neural computer consists of a neural network where quantum processes are supported*".

These statements can be considered as implicit definitions of a quantum neural network: a connectionist system characterized by a wave function, or a neural network that supports quantum processing (here it is necessary to remember why such non-trivial elements as artificial neurons usually have to be connected in some kind of network - this is due to their non-universality: a separate artificial neuron with a threshold transfer function (McCulloch-Pitts neurons [7]) cannot implement an arbitrary Boolean function, as a consequence, it is necessary (and sufficient) to build at least a two-layer neural network. A single linear or nonlinear neuron also cannot approximate a continuous function of many variables - again, you should at least use a similar two-layer network. If you imagine the concept of a quantum neuron, the need to build networks from them is not so obvious *a priori*. Note that speaking about the wave function describing a quantum neural network, we used the language of the canonical Copenhagen interpretation of quantum mechanics. The presented definition of quantum neural networks was used and refined by Tammi Menneer and Ajit Narayanan [8,9]. They presented two approaches to creating a new neural network model inspired by quantum concepts:

> *"A neural network approach strongly inspired by quantum concepts regards each pattern in training set as a particle which is processed by a number of distinct neural networks (probably homogeneous but not necessarily so) in different universes. A weakly inspired approach regards each pattern in training set as a particle which is processed in its own universe and no other"* [8,9].

It is noteworthy that the authors of [8] explicitly use the many worlds interpretation of quantum mechanics [10–12], a proponent of which David Deutsch is one of the fathers of quantum computing [13]. So, the authors suggested that a quantum neural network is a superposition of classical neural networks, each of which exists in its own world. Recall that, according to Kak's formulation, such a quantum neural network is obviously described by a single wave function.

Although all classical neural networks in every world can have any structure and consist of neurons of any type with different transfer functions (nonlinear in general, but not necessarily nonlinear), Tammy Menneer, in her dissertation [14], took a weakly inspired approach (assuming that in each world, the network learns only one pattern from the dataset). Note that Narayanan and Moore used a similar interpretation of quantum mechanics to propose a quantum-inspired genetic algorithm [15]. We also note that the author of the first known quantum algorithm, Peter Shor, characterized Copenhagen and many worlds interpretations as useful for various situations [16]:

> *"There are some times when thinking about quantum mechanics using the Copenhagen interpretation will help you figure things out, and there are other times when the many-worlds interpretation is more useful for figuring things out. Since these two interpretations give the exact same predictions, it doesn't matter which one you use. So you should use whichever gives you the best intuition for the problem you're working on."*

Further proposals clearly used *strongly inspired* approaches and the papers of Chrisley [17] and Behrman and Steck with colleagues [18-19] were among the most thought-provoking. Unlike Narayanan and Menneer model, they used canonical quantum mechanics. It has two forms – Hamiltonian's and Lagrangian's [20]. The Hamiltonian's form consisting of the *linear* Schrödinger equation and the *nonlinear* Born law [21], describing the collapse of the wave function during measurement, is a convenient tool for calculating the spectra of quantum systems, such as atoms, nuclei, etc. The Langrangian's form – the Feynman path integral formalism [22] – is suitable for calculating the amplitudes of various quantum processes, such as scattering, reactions, etc. Behrman et al [18] applied Feynman path integral as a starting point for constructing their model of *temporal quantum neural network,* by considering time-intermediate quantum states as quantum neurons belonging to different network's layers. The authors emphasized that the nonlinearity typical for artificial neural models is *already present* in the path integral both in the form of an exponent and also in the nonlinear form of kinetic energy, which is a part of the system's Lagrangian. The model proposed by Chrisley [17] considers a system (similar to the classical two-slit model) consisting of a photon source that illuminates a screen containing many variable-position slits, some of which can encode inputs to a quantum neural network, and others - the values of the trained weight parameters.

The ideas embodied in these two models were used in [23, 24] when constructing a system in which media with different refraction indices were located between several multiple-slit barriers defining the input vector of analog quantities, and the position of the slits on the barriers were used to vary the weights, the values of which were determined by the geometric paths of photons between the slits located on adjacent barriers. A similar scheme (albeit without refractive media between screens), called the "computation of quantum paths", was recently proposed and investigated in [25, 26]. The authors of [23] called this system a *quantum neuron* and argued that there is no need to connect such systems into any network, because this quantum neuron is already capable of approximating any continuous function [24]. Now we believe that this statement is not enough to clarify the concept of a quantum neural network (such a network has not been considered), but both this model and the Behrman's model lie half a step before such a definition.

Note that the system proposed earlier in [23,24] corresponds to the definition of a neural network approach, *strongly inspired* by the quantum concepts formulated by Narayanan and Menneer, since each input here is processed by quantum neurons, which corresponds to the path described by exponent factor – the path amplitude. At the beginning of the study of quantum neural computing, the amount of work in this field was rather modest. Nevertheless, these works used different formulation of quantum mechanics and already utilized nonlinearity inherent to QM, even without reference to the mechanism of collapse. This early stable period in the field of quantum neural networks changed after the invention of new ways to train *deep neural networks* (having dozens of intermediate levels) [27]. The remarkable success in the application of these systems (connected with the invention of rectified linear neurons, as well as the availability of large unclassified data and GPUs) in many fields revived interest in creating quantum analogs of neural networks. If from 1995 until 2005 the number of papers on this topic was about 50, then since 2005 this number has increased by an order of magnitude [1].

## 2. The quest for nonlinearity

Further research in this area can be characterized as attempts to combine not neural networks with quantum mechanics, but neural networks with quantum informatics [28]. The first neural networks actually worked with binary inputs or Boolean variables, which made it possible to establish their correspondence to logic (in fact, the same logical function can be implemented using many different neural networks, but a two-layer architecture is enough for such an implementation to be simple and universal ). Of course, such neural networks can implement the basic operations of computers, so their connection with the concept of bits is obvious. The real history of computers has shown that a different basis was used to build them. The fact is that in the future, an artificial neural network worked not with bits, but with real, or even complex data. On the other hand, generalizing bits to the quantum domain leads to the concept of a qubit, and since quantum computers use qubits as the basic building blocks, many researchers have tried to combine neural networks and qubits with many of the difficulties involved. As mentioned in the review by Maria Schuld et al. [29]:

> "*The majority of proposals for QNN models are consequently based on the idea of a qubit neuron (or 'quron' as we suggest to name it), and theoretically construct neurons as two-level quantum systems*".

This approach is associated with the need to solve both some natural problems, such as the representation of classical information in quantum systems (for example, phase and amplitude coding), and some fictitious problem, such as solving the "*problem of combining the nonlinear dissipative dynamics of neural computations and linear unitary dynamics of quantum computing*" [29]. We will not discuss all the very clever and complex approaches to the last question, such as the use of a clearly nonlinear phenomenon of collapse of the wave function during measurement [30], nonlinear transformation of classical data into quantum data, repeat to success (RUS) approach [31-33] (which also uses measurements), the use of single-mode Kerr interactions [34], etc. - many published reviews are devoted to these schemes [29, 35-38]. Note that both parts of this claimed conflict are mis-delineated. Linear artificial neurons, although scarce, are an important part of a universal approximating neural network, and on the other hand, quantum mechanics does have a non-linear element beyond Born collapse. Note that only a few remarkable works really declare the nonlinearity inherent in quantum neural networks [26, 39-40]. Note also that some articles do not consider nonlinearity at all, for example, some authors suggest that "*a truly quantum analogue of classical neurons*" can be defined as "*an arbitrary unitary operator with m input qubits and n output qubits*" [41]. Moreover, this problem does not exist if only the existence of training parameters is sufficient to classify the system as a "*quantum neural network*" [42]:

> "*Quantum neural networks are a subclass of variational quantum algorithms, comprising of quantum circuits that contain parameterized gate operations.*"

Such an algorithm was introduced in [43]. These circuits also do not consider the problem of nonlinearity [44].

## 3. Brief history of classical artificial neural networks

The McCulloch-Pitts neural network model was proposed in 1943 and inspired by natural neural systems [45]. The authors suggested that artificial neurons can use a binary code corresponding to the transmission (or not transmission) of signals along their output branching axon. These signals can reach other neurons, which adds the influence of about tens of thousands of them through various synaptic connections that can act together to excite or suppress neuron activity. McCulloch and Pitts suggested that the response of a neuron is non-linear, described by the Heaviside threshold function (hence the other name for them - threshold neurons). The most notable result was that these binary characteristics of threshold neurons were interpreted as Boolean variables, and the network of neurons as a system capable of implementing Boolean functions. The main result is associated with the establishment of the universality of such networks - a simple two-layer network can implement an arbitrary Boolean function, with the correct choice of the values of the synaptic weights of neurons. Thus, in a certain sense, the equivalence of threshold neural networks and Boolean algebra was shown ("*anything that is sufficiently finite*" that could be logically represented can be performed by a neural network [45]). Earlier, in 1937, Claude Shannon showed how any logic function can be implemented using electronic switching circuits [46].

This work deeply influenced the development of electronic computers in the US and UK over the next decades. Was it possible to choose threshold neural networks as an element base for classical computers? Of course, "*yes*", but history took a different path.

Can we "see" a universal two-level network in the brain? In a sense, "yes", if we differentiate sensory neurons (they are usually modeled not as neurons, but as inputs) and output motor neurons and attach all other intermediate neurons (interneurons) on the "hidden" layer. Unfortunately, this is a big stretch, because this layer is too complex and huge and almost coincides with the entire brain).

However, we can find a system very much like a universal neural network in a very unexpected place - the US electoral system. In full accordance with the two layers of the universal neural network, a two-step process is used. The first hidden layer corresponds to elections in all States (thus, the number of hidden neurons is the number of States plus the District of Columbia equal to 51). The voting results form a threshold decision (for example, the victory of the Democrats corresponds to the state of the binary spin output of the neuron equal to −1 otherwise +1). The output neuron then multiplies (weights) these values by the number of electors defined for each State, and the results are summed up. The final decision of the electoral college is also of a threshold nature − the conservatives win if this amount is positive.

The last example has the sole purpose of illustrating that sometimes we can "*see*" something like a universal neural network in rather strange places. This situation arose again when continuous output neurons described by a differentiable transfer function were introduced. This made it possible not only to interpret the output values as the probability of the input patterns belonging to a given class, but also to find a very effective algorithm for learning them − the *backpropagation error method* [47-50].

The main question again was what kind of neural architecture would be universal to represent an arbitrary classification of any dataset. Undoubtedly, this architecture must be associated with a fundamental mathematical structure. The first place researchers saw this new neural architecture was the solution to Gilbert's 13th problem found by Kolmogorov [51] and Arnold [52]. The theorem they proved states that any function of many variables can be constructed as a composition of functions of only one variable. Here, many variables are treated as inputs to neurons, and one variable is treated as a weighted sum of inputs. It is noteworthy that the structure of this composition resembled a 3-layer neural network. But it turned out that the functions of one variable in the Kolmogorov-Arnold theorem are very crazy (not differentiable at all points) and cannot be considered as a reasonable nonlinear transformation produced by neurons.

Later it was recognized that the universal network should not accurately represent the value of a function of many variables, but only approximate this value with any accuracy. Then the structure of the universal network was found (seen) in the Stone – Weierstrass theorem [53]. Remarkably, its architecture again turns out to be two-layer, one output neuron of which is generally linear (in the case of a probabilistic interpretation, it can be logistic – this is easily seen from Bayes' theorem) [54-56]. Then it was found that the nonlinear transfer function of hidden neurons can have a rather general form, for example, sigmoid and even periodic (in the general case, an arbitrary limited and variable activation function [57]). In addition, neurons can have complex outputs (such networks can outperform networks with real outputs [58]).

These results take us far from threshold neurons with binary inputs and outputs interpreted as Boolean variables. In 1982, Hopfield saw that such variables could be thought of as "*spins*", and also saw a very fruitful analogy between fully interconnected threshold neural networks and spin glasses [59], which allowed artificial neural networks to become the legal section of *Physical Review B*. That is, Hopfield saw his the famous *attractor neural network* (where all neurons are simultaneously input, hidden and output) in condensed matter physics (note that the Hopfield network, as well as threshold neural networks and an approximating network, have universality properties, which are essentially an emergent property of complex systems of interacting nonlinear elements). Hopfield's model attracted physicists to the field of neural networks, and the application of complex methods of theoretical physics has become a new amazing and very powerful tool in the field in which neurophysiologists, mathematicians, computer scientists, etc. previously worked. In fact, the field of neural networks has become part of physics [60–61], and we can assume that, to paraphrase the medical proverb "*Becoming a physics, it will forever remain physics*" [62].

Remind that classical computers can be built using the architecture of threshold neurons, in which logical variables should be interpreted as the central concept of information theory - bits. Quantum informatics arises from classical information by generalizing this concept to qubits. Many attempts to create a quantum version of neural networks follow this path, not counting the fact that the artificial neural network almost forgot the bits and divorced with classical computers many years ago. But the remarkable characterization of the modern mainstream ignores this fact, as clearly seen in the title of a recent article, "*The Future of Computing: Bits + Neurons + Qubits*" [63]. The first model of this type, rooted in quantum informatics, the *quantum perceptron*, was proposed by Altaysky [64]. Numerous other models have also been proposed in recent years and discussed in [29, 34, 37-38]. We will argue that a quantum generalization of neural networks yielding a universal quantum neural network cannot be obtained from classical or quantum computing, classical or quantum information, or biology, but can be seen in quantum mechanics itself.

## 4. Quantum neural network and path integral

Our approach is based on the Feynman path integral formulation of quantum mechanics and is similar to that used by Berman et al. [18]. However, instead of introducing neurons for every sampling point in time [18], we consider every path the quantum system can follow like a quantum neuron. A path integral formulation is a description in quantum mechanics that generalizes how classical mechanics works. It replaces the classical notion of a single unique classical trajectory for a system with a sum or functional integral over an infinite number of possible quantum mechanical trajectories to calculate the quantum amplitude.

Here we will use the sum of histories, each of which can be thought of as the classic path that a quantum system can take. In this case, the amplitude of the system of transition from the initial point of space-time to the final one is given by the sum of complex exponential factors along all possible paths (1), where the phase of each exponent is defined as the action $S$ or the time integral of the Lagrange function (2).

$$A = \sum_{\{path\}} e^{i\frac{S_{path}}{\hbar}} \qquad (1)$$

$$A = \sum_{\{path\}} exp\left\{\frac{i}{\hbar} \int_0^T L(t)dt\right\} \qquad (2)$$

Let us compare the expression for the amplitude (2) with the output of the classical universal approximating two-layer neural network (3). This network has a single linear output neuron with weights $W_h$ and nonlinear hidden neurons with transfer function $(.)$. Expression (2) formally coincides with (3) if we put all the connection weights of the output neuron equal to 1 and choose exponent as transfer function. The arguments of the exponents are integrals over time that correspond to the activation of hidden neurons in (3) defined as the weighted sum of the inputs $x_j$ to these neurons.

$$y = \sum_h W_h f\left(\sum_j w_{j \to h} x_j\right) \qquad (3)$$

Formal equivalence becomes more obvious if we consider multiple-barrier multiple-slit systems shown in Fig.1. In this case, the phase of the exponents is transformed into weighted sums of refraction indexes describing inter-barrier media, where the geometric paths of photons with the factor $2\pi/\lambda$ can be considered as weights. Thus, these refraction indexes form a real-valued input vector.

$$A = \sum_{\{path\}} exp\left\{i \frac{2\pi}{\lambda} \sum_j l_{j,path} n_j\right\} \qquad (4)$$

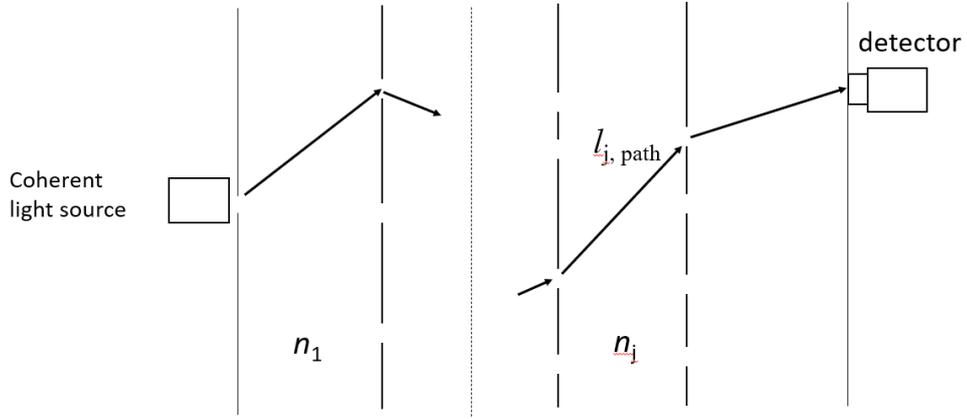

**Figure 1.** Multiple-barrier multiple-slit system that implements quantum neural network. The probability amplitude of detecting a photon emitted by a coherent source is the sum (output of a linear neuron) of paths amplitudes (output of hidden neurons) defined by the exponent of action (activation of hidden neurons) which is the sum of optical paths (product of the geometrical inter-barrier path $l_{j,path}$ and the corresponding refraction index, $n_j$). Geometrical inter-barrier paths imitate the weights of hidden neurons, the values of which can be controlled by the positions of slits. The refraction index vector defines the real-valued input vector processed by this network. This system effectively has the two-layer neural architecture.

Note that equation (4) can be derived from (2) by setting $L(t) = T - V$, where for a photon in free space $T = \hbar\omega$, $V = 0$, $dt = \frac{ds}{v} = n\frac{ds}{c}$ [65]. In the more general case of a nonrelativistic particle of mass $m$ moving in a one-dimensional potential $V(x)$, it is possible to use the analogy between optics and quantum mechanics [66] and find effective refraction index corresponding to a given potential

$$n^2(x) = 1 - \frac{2m}{\pi^2 k^2} V(x), \ k = \frac{2\pi}{\lambda} \qquad (5)$$

Then an expression for the amplitude similar to (4) can be obtained when the real function $n(x)$ is an input with an infinite number of components, and sum representing the action in (4) in the general case is transformed into an integral. We can conclude, that the amplitude of the quantum process, expressed as the sum of the paths, is simply the output of some two-layer network, where the path amplitudes determine the complex-valued outputs of neurons, and the number of inputs is determined by the number of inter-barrier spaces. The number of hidden neurons is equal to the number of all possible paths, and the output neuron weights the outputs of the hidden neurons with the same weights equal to one.

Note again that, in general case, the number of inputs to hidden neurons can be determined after sampling the time interval in the integral representing the action, and can be very large (see [18-19]), but, as will be demonstrated below in some interesting applications, this action can simply be obtained analytically. But more importantly, the number of hidden neurons themselves, determined by the number of paths, can be enormous. This means that a universal quantum neural network can be characterized as *wide neural network* (WNN) in contrast to the widely used *deep neural network* (DNN) which can have dozens of hidden layers. Note that a universal two-layer network of McCulloch-Pitts neurons capable of realizing any Boolean function is also a wide network – for a function of $d$ variables it must have an exponential number, $2^d$, of neurons in the hidden layer.

It can be concluded that the output of a two-layer quantum neural network corresponds to the quantum amplitude of an arbitrary quantum process. Each hidden neuron of a two-layer quantum neural network corresponds to a separate path in the sum over histories. A universal quantum neural network capable of representing any quantum amplitude has a linear output quantum neuron, as in universal approximating classical neural networks. It is noteworthy that three universal neural networks (two classical and one quantum) have a similar two-layer architecture.

## 5. The Universe as universal quantum neural network

The general form of the sum of histories clearly shows, that in any case it can be considered as a two-layer quantum neural network. Of course, this relation becomes more obvious if we can represent the action as the activation of neurons in hidden layers. This can be done, for example, by using a optic-mechanical analogy to describe corresponding physical system. For gravity, this approach was proposed by Eddington, who showed that the gravitational field can be considered as an optical medium described by effective refraction indices [67]. This approach was later used by different authors. For example, Evans et al [68] demonstrated, that for the Schwarzschild metric for the mass $M$ of the central gravitating body and central gravitational field with singularity at $r = 0$:

$$d\tau^2 = \left(1 - \frac{2MG}{r}\right)dt^2 - \left(\frac{1}{1-\frac{2MG}{r}}\right)dr^2 - r^2 d\Omega^2 \tag{6}$$

(where the speed of light is $c = 1$, $G$ – is the universal gravitational constant) the effective refraction index at $r > M/2$ for both light and also for massive particles has the form:

$$n(r) = \left(1 + \frac{MG}{2r}\right)^3 \left(\frac{1}{1-\frac{MG}{2r}}\right) dr^2. \tag{7}$$

In general, one can avoid optical the optical analogy and directly use the form of action, considering it as an analog of the activation of hidden quantum neurons (paths). For example, for the particle having the mass $m$ and moving radially (here $d\Omega^2 = 0$), for example, from the Black Hole of mass $M$ from the space-time point 1 (out of the Black Hole) to point 2 the action $S$ will take the form:

$$S = -m \int_1^2 dt \sqrt{\left(1 - \frac{R_S}{r(t)}\right) - \frac{1}{1-\frac{R_S}{r(t)}}\dot{r}^2}, \qquad (8)$$

where $R_S = \frac{2MG}{c^2}$ is the Schwarzschild radius of the Black Hole. Taking into account that the energy of the particle

$$E = \frac{mc^2\left(1-\frac{R_S}{r(t)}\right)}{\sqrt{\left(1-\frac{R_S}{r(t)}\right)-\frac{1}{1-\frac{R_S}{r(t)}}\dot{r}^2}} \qquad (9)$$

is conserved we can express $\dot{r}^2$ from the last expression and put it into the expression of action

$$S = -mc^2 \int_1^2 dt \left(\frac{mc^2}{E}\right)^2 \left(1 - \frac{R_S}{r(t)}\right). \qquad (10)$$

It can be seen that $S$ represents the activation of a neuron with single input

$$-mc^2 \left(\frac{mc^2}{E}\right)^2 \qquad (11)$$

and synaptic weight

$$\int_1^2 dt \left(1 - \frac{R_S}{r(t)}\right), \qquad (12)$$

where $r(t) > R_S$. Note, that real neurons do integrate inputs over time and this integral determines their activation. For artificial neurons, their input data can be highly heterogeneous (representing completely different features of the presented object, for example, weight, height, race and religion of a person). In physics this integral (action) can be taken and factored by the geometric parameter corresponding to the weight and by the matter field parameter corresponding to the input (compare with path length and refraction index in optical model). An analysis of expressions for the action in various metrics can be used to make sure that a quantum neural analogy can be seen in models of quantum cosmology, which describes the origin and early stages of the evolution of the Universe as a whole.

There are two main approaches to the evolution of the Universe in quantum Cosmology. One of them postulate the wave function of the Universe which obeys the Wheeler- deWitt equation. The other is based on the use of a path integral, which describes the evolution of the Universe as the sum of paths through various possible geometries and matter fields. There are many approaches using different form of action in these studies. Nevertheless, one can guess that the amplitude of the evolution of the Universe really has the architecture of two-layer quantum neural network.

$$S = \int d^4x \sqrt{-g} \left[\frac{c^4}{16\pi G} R + L_M\right]. \qquad (13)$$

Here we easily recognized the term in the brackets as matter field input $L_M$, and geometric parameters $\frac{c^4}{16\pi G} R$ and $\sqrt{-g}$ define the threshold and weights, correspondingly.

In general, the form associated with matter fields is quite complex and does not obviously resemble the simple form of neural activation as in the case of the multiple-barrier multiple-slit model. We know that historically various forms of meaningful transformation functions have been considered – from threshold to rectified linear – but the form of activation is usually viewed as the dot product of input and weight vectors. Of course, it is not unnatural to consider other forms of activation to generalize the form of the network, which is still two-layer. Real neurons are activated by integrating over time inputs from many synaptic junctions. Thus, the form of *activation* as *action* (note the interesting similarity of these words) which is Lagrangian integral over time, seems to be quite general. The kinetic energy, which is part of Lagrangian seems to give a different type of nonlinearity along with an exponential amplitude dependence. Also, integrating over the time interval is like summing over an infinite number of inputs. This circumstance is not a problem, and we will give several examples when such an integral takes a simple form.

In cosmology, various cases of the Friedmann equation can be associated with the simple problem of particles (rocks, stones) thrown upwards from the Earth's surface. In this case, the Lagrangian takes the form:

$$L = \frac{m}{2}\dot{r}^2 + \frac{GMm}{r}. \tag{14}$$

Newton's equation

$$m\ddot{r} = -\frac{GMm}{r^2} \tag{15}$$

can be transform into the equation

$$vdv = -GM\frac{dr}{r^2}, \tag{16}$$

where $v = \dot{r}$. By integrating its right-hand side over $r$ from the Earth's radius $R_E$ to the current radius $R$ and its left-hand side from the escape velocity

$$v_{escape}(R_E) = \sqrt{\frac{2GM}{R_E}} \tag{17}$$

to the velocity at radius $R$, $v(R)$, we find

$$v(R) = \sqrt{\frac{2GM}{R}}. \tag{18}$$

Taking this expression into account, calculate the action as

$$S = \int_{R_E}^{R} \frac{dr}{v(r)} \left(\frac{mv(r)^2}{2} + \frac{GMm}{r}\right) = m\int_{R_E}^{R} dr \sqrt{\frac{2GM}{r}} = m\sqrt{8GM}(\sqrt{R} - \sqrt{R_E}). \tag{19}$$

It is clear that action (activation) is a direct product of material (input-like) and geometrical (weight-like) factors similar to the activation of a neuron with one single input. The following example shows that in cosmological models, an infinite number of inputs can also be equivalent to one.

*Cosmology of de Sitter space*

This analogy can be used in cosmology models (not necessarily quantum ones). Consider the Friedman equation

$$\left(\frac{\dot{a}}{a}\right)^2 = \Lambda - \frac{1}{a^2} \tag{20}$$

which describes a positively curved spherical Universe. Here $a(t)$ is the scale factor and $\Lambda$ is the cosmological factor. Denoting $\lambda^2 \stackrel{\text{def}}{=} \Lambda$, we transform it into

$$\dot{a}^2 - \lambda^2 a^2 = -1. \tag{21}$$

Solving this equation we get

$$a(t) = \frac{1}{2\lambda}\left(e^{\lambda t} + e^{-\lambda t}\right). \tag{22}$$

Considering $a$ as particle's line coordinate, $\dot{a}^2$ as the particle's kinetic energy, $-\lambda^2 a^2$ as it's potential energy and their sum, $-1$ (right-hand side of equation) as the total energy we get that Lagrangian

$$L = T - U = E - 2U = -1 + 2\lambda^2 a^2. \tag{23}$$

Taking into account the solution for $a(t)$ (22), it is easy to calculate the value of the action by integrating the Lagrangian over time from $t = 0$ up to $t = T_{now}$:

$$S = \int_0^{T_{now}} dt L(t) = H(T_{now})a^2(T_{now}) = tanh(\sqrt{\Lambda}T_{now})a^2(T_{now}). \tag{24}$$

We see that the action is also the result of multiplying the value of input-like physical parameter (a function of the cosmological parameter associated with dark energy) and the weight-like geometric parameter – squared scale factor taken at the current time $T_{now}$. Formally, this is again is equivalent to the activation of neuron which has the only input.

*Action for Inflation*
Another example can be given by the inflation period, when the Lagrangian of the scalar inflation field $\varphi$ with high but steepest potential field $V(\varphi)$ has the form

$$L(t) = \left(\frac{\dot{\varphi}^2}{2} - V(\varphi)\right)a^3(t) \cong -V(\varphi)a^3(t) \tag{25}$$

(the kinetic term of the scalar inflation field is negligible). In this case the approximate Friedman equation

$$H^2 = \left(\frac{\dot{a}}{a}\right)^2 = \frac{8\pi G}{3}V(\varphi) \tag{26}$$

has a solution

$$a(t) = e^{Ht}, \tag{27}$$

where the Hubble constant

$$H(\varphi) = \sqrt{\frac{8\pi G}{3}V(\varphi)} \tag{28}$$

and the field potential $V$ can be considered constant. Consequently, the action for the Universe during the inflation period is

$$S = \int_0^T dt L(t) \cong -\frac{V}{3H}e^{3HT} = -\frac{V}{3H}a^3(T) = -\frac{1}{3}\sqrt{\frac{3V}{8\pi G}}a^3(T). \tag{29}$$

Again, the action can be viewed as the product of the geometric weighting factor $a^3(T)$ – the volume of the inflated Universe and the input energy factor determined by the potential of the inflaton field.

So, although time-points seem to present an infinite number of inputs to hidden non-linear neurons (see Behrman [18]), the integration of the Lagrangian gives, as in the previous cases, an activation similar to the activation of one-weight one-input neuron. Note, that real neurons, such as very important and unusual spindle (*von Economo*) neurons [69], whose architecture resembles a single (at least single-dendrite) input neuron exhibit a complex activation pattern based on time integration of distant from the soma synapses.

Can we now consider the Universe as a universal quantum neural network? Or, conversely, as Seth Lloyd says: "*The universe is a quantum computer*" [3]. Interestingly, Wharton objected to Lloyd's statement, and his argument used the Lagrangian approach to quantum mechanics and path integrals [70]. He concluded that:

"*Constructing a complete theory built upon the Lagrangian Schema is a vast project, one that has barely even begun. It is these models, the balance of the evidence suggests, that have a chance of representing how our universe really works. Not as we humans solve problems, not as a computer, but as something far grander*" [70].

Vanchurin recently suggested that the Universe as a whole could be a neural network with ordinary nonlinear hidden neurons, but not a quantum neural network [71].

We assume that. at least for the problems considered in quantum cosmology, the answer to our question can be positive: <u>The Universe can be considered as universal two-layer quantum neural networks.</u>

## 6. Conclusions

In this paper we tried to argue the following:
1) The field of *quantum neural networks* is the field of *physics* not *machine learning*.
2) A universal quantum neural network represents the amplitude of an arbitrary quantum process
3) The structure of a universal quantum neural network – a two-layer network – is identical to the structure of universal classical neural networks for the implementation of Boolean function and the approximation of continuous functions.
4) The universe as a whole can be considered as quantum neural network.

## Acknowledgements

The author thanks Andrey G. Khromov for his attention to the work and its constructive criticism, and also expresses heartfelt gratitude to Yuri I. Ozhigov, who encouraged the preparing this material.